\documentstyle[11pt,psfig,latexsym]{article}

\begin{document}

\title{Energy Levels of Quasiperiodic Hamiltonians, Spectral 
  Unfolding, and  Random Matrix Theory}
\author{M.~Schreiber$\,^1$, U.~Grimm,$^1$
  R.~A.~R\"{o}mer,$^1$ and \\J.~X.~Zhong$\,^{1,2}$\\
$^{1}$Institut f\"{u}r Physik, Technische Universit\"{a}t, 
\\D-09107 Chemnitz, Germany\\
$^{2}$Department of Physics, Xiangtan University,
\\Xiangtan 411105, P.~R.~China}   

\maketitle

\begin{abstract}
  We consider a tight-binding Hamiltonian defined on the quasiperiodic
  Ammann-Beenker tiling. Although the density of states (DOS) is
  rather spiky, the integrated DOS (IDOS) is quite smooth and can be
  used to perform spectral unfolding. The effect of unfolding on the
  integrated level-spacing distribution is investigated for various
  parts of the spectrum which show different behaviour of the DOS. For
  energy intervals with approximately constant DOS, we find good
  agreement with the distribution of the Gaussian orthogonal random
  matrix ensemble (GOE) even without unfolding. For energy ranges with
  fluctuating DOS, we observe deviations from the GOE result. After
  unfolding, we always recover the GOE distribution.
\end{abstract}

\begin{figure}[t]
\centerline{\psfig{figure=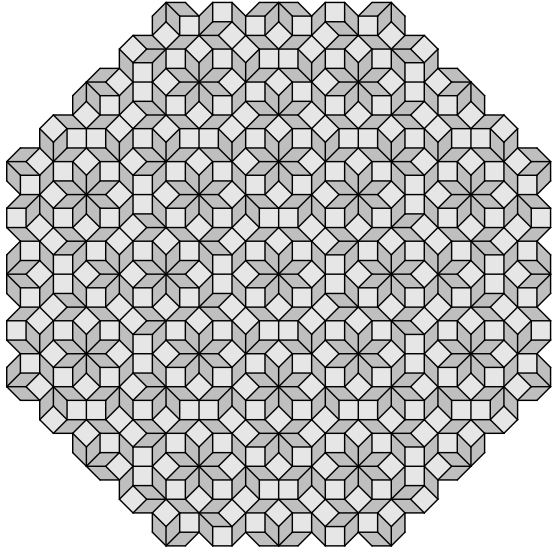,width=0.6\textwidth}}
  \caption{
    Octagonal cluster of the Ammann-Beenker tiling with $833$
    vertices and an exact $D_8$ symmetry around the central vertex.}
\label{fig1}
\end{figure}

In two recent papers \cite{zgrs,sgrz}, we investigated the
energy-level statistics of two-dimensional quasiperiodic Hamiltonians,
concentrating on the case of the eight-fold Ammann-Beenker tiling
\cite{AB} shown in Fig.~\ref{fig1}. The Hamiltonian contains solely
constant hopping elements along the edges of the tiles in
Fig.~\ref{fig1}. Numerical results suggest that typical eigenstates of
the model are multifractal \cite{RSS}. In \cite{zgrs,sgrz}, we numerically
calculated the level-spacing distribution $P(s)$ and the $\Delta_3$
and $\Sigma_2$ statistics \cite{MH}, and found perfect agreement with
the results for the GOE. One ingredient of the calculation was the
so-called unfolding procedure, explained below, which corrects for the
fluctuations in the DOS of the model Hamiltonian. It is well known
that, for a spectrum with non-constant DOS, it is necessary to unfold
the spectrum in order to extract universal level statistics which can
then be compared to the results of random matrix theory \cite {MH}.

\begin{figure}[t]
\centerline{\psfig{figure=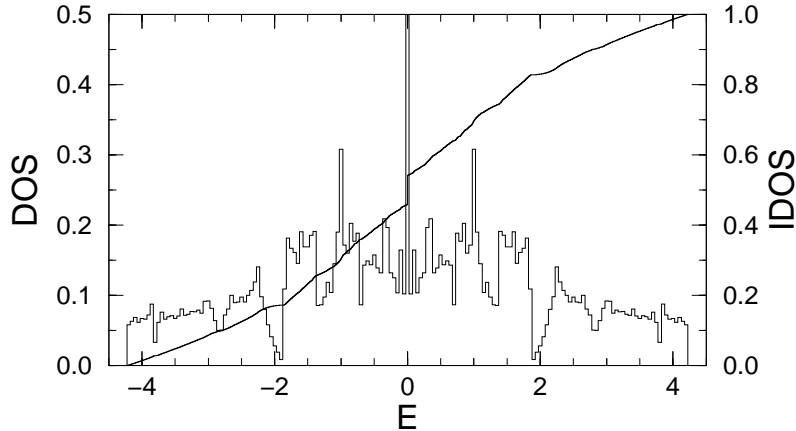,width=0.85\textwidth}}
  \caption{DOS and IDOS for a $D_8$-symmetric patch with
  $157369$ vertices.}
\label{fig2}
\end{figure}

In the present paper, we trace the question how crucial the unfolding
procedure affects the universal results, restricting ourselves to the
integrated level-spacing distribution (ILSD)
$I(s)=\int_{s}^{\infty}P(s')\, ds'$. In previous studies of the model
\cite{BS,PJ,ZY}, spectral unfolding had not been used, the main reason
being that the DOS is rather spiky as shown in Fig.~\ref{fig2}. In
\cite{P}, it was claimed that the level-spacing distribution (LSD)
depends crucially on the unfolding, yielding results ranging from
log-normal to GOE behaviour. In fact, if one considers the IDOS which
is also shown in Fig.~\ref{fig2}, one finds that it is rather smooth,
apart from a couple of small gaps and a jump at energy $E=0$ caused by
the existence of 13077 degenerate states in the band center
\cite{zgrs}. These states can be neglected for the level-spacing
distribution since they only contribute to $P(0)$. Therefore, we
use the IDOS to unfold the energy spectra.

The purpose of the spectral unfolding is to transfer a spectrum with a
non-constant DOS to the another with a unit DOS or a linear IDOS. In
order to achieve this, one defines the unfolded levels as $e_i=N_{\rm
  av}(E_i)$, where $E_i$ is the $i$-th energy level, $N_{\rm av}(E_i)$
the smoothened IDOS \cite {MH}. Level spacings are thus given by
$s_i=e_{i+1}-e_i$.  An effective way to calculate the $N_{\rm
  av}(E_i)$ is to fit the IDOS to cubic splines by choosing a sequence
of energy levels $(E_1, E_{m+1}, E_{2m+1}, \ldots)$ from the energy
spectrum $(E_1,E_2,\ldots,E_n)$.  A meaningful statistics should be
independent of the unfolding parameter $m$.  Evidently, the value of
$m$, which does not change the statistics, depends on the total number
of levels $n$ and the structure of the DOS.  The typical value for the 
Anderson Hamiltonian is around $100$ \cite {HS, ZK}.
 
The eight-fold Ammann Beenker tiling shown in Fig.~1 has the full
$D_8$ symmetry of the regular octagon, hence the Hamiltonian matrix
splits into ten blocks according to the irreducible representations of
the dihedral group $D_8$, resulting in seven different independent
subspectra as there are three pairs of identical spectra. In order to
obtain the underlying universal LSD, one should consider the
irreducible subspectra seperately \cite {zgrs, MH}. In the following,
we demonstrate our results by concentrating on one of the largest
subspectra of the Ammann Beenker $D_8$-symmetric tiling of $157369$
vertices. We note that the derived conclusions hold for all the
subspectra.  Fig.~3 shows level spacing distributions for energy
levels in various parts of the spectrum without and with unfolding. It
is easy to see from Fig.~3(a) that energy levels with an approximately
constant DOS (for instance, $3.2\le E\le 3.3$) exhibit GOE behaviour
even without unfolding.  However, for energy ranges with fluctuating
DOS, $I(s)$ apparently deviates from the $I_{\rm GOE}(s)$ if we do not
perform the unfolding.  For instance, $I(s)$ for the non-unfolded
whole spectrum is close to the log-normal distribution although the
log-normal distribution is not the generic level statistics of the
quasiperiodic Hamiltonian.  Analyzing the non-unfolded spectrum in the
range with the largest fluctuations, $0.5\le E\le 1.5$, one can
clearly see that $I(s)$ is neither log-normal nor GOE.  From
Fig.~3(b), we can see that, after the unfolding, $I(s)$ for different
energy intervals show a very good agreement with the GOE result.

\begin{figure}[t]
  \centerline{\psfig{figure=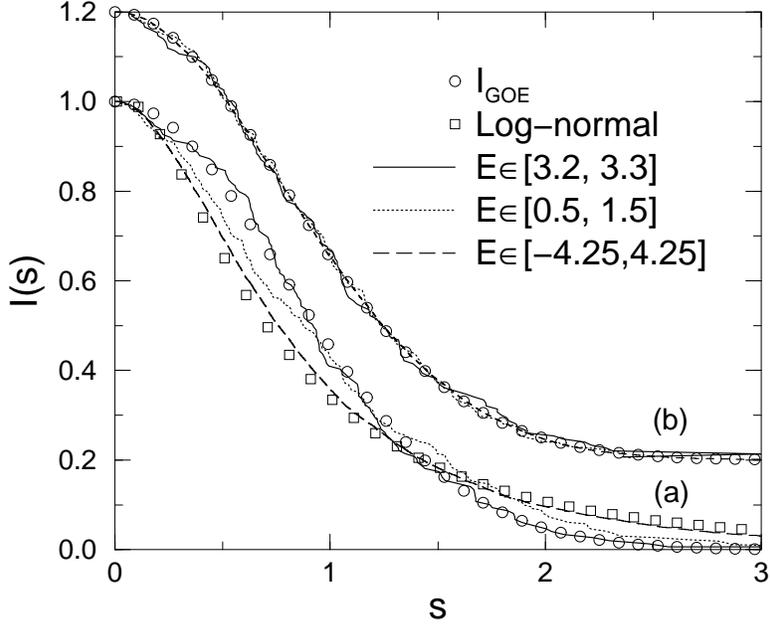,width=0.8\textwidth}}
\caption{ILSD $I(s)$ obtained (a) without unfolding  
  and (b) with unfolding for various parts of the spectrum of one
  sector of the $D_8$-symmetric patch: whole spectrum (dashed line),
  $0.5\le E\le 1.5$ (dotted line), and $3.2\le E\le 3.3$ (solid line).
  Circles and boxes denote $I_{\rm GOE}(s)$ and the log-normal
  distribution, respectively. $I(s)$ in (b) has been shifted by $0.2$ 
  for clarity.}
\label{fig3}
\end{figure}

In our calculation, we find that there exists a sequence of $m$
ranging from $4$ to $30$, which gives the same statistics described by
$I_{\rm GOE}(s)$, within our numerical precision.  In Fig.~4, we
illustrate $I(s)$ obtained with $m=5, 10$, $20$, and $100$. One can
see that there is only a small deviation from the $I_{\rm GOE}(s)$
even for large $m$ up to $100$.  We emphasize that even for such a
large $m$, $I(s)$ is still far from the so-called ``semi-Poisson''       
intermediate statistics $I_{\rm SP}(s)=(2s+1)e^{-2s}$ \cite {BG},
which is supposed to be valid to describe the level statistics at the
metal-insulator transition with multifractal eigenstates in the
three-dimensional Anderson model of localization.

In summary, we have shown that although the DOS of the quasiperiodic
tight-binding Hamiltonian defined on the Ammann-Beenker tiling
exhibits very spiky structures, the IDOS is in fact rather smooth  and
can be used to perform spectral unfolding. We demonstrated that the
level spacing distribution is independent of the unfolding parameter
and is well described by the GOE distribution. We also studied
the level statistics for energy spectra without unfolding and found
that, for energy intervals with approximately constant DOS, $I(s)$
agrees with the GOE distribution; for energy ranges with fluctuating
DOS, $I(s)$ varies with different energy intervals and the log-normal
distribution found in a previous calculation \cite {PJ} is not
generic.

\begin{figure}[t]
\centerline{\psfig{figure=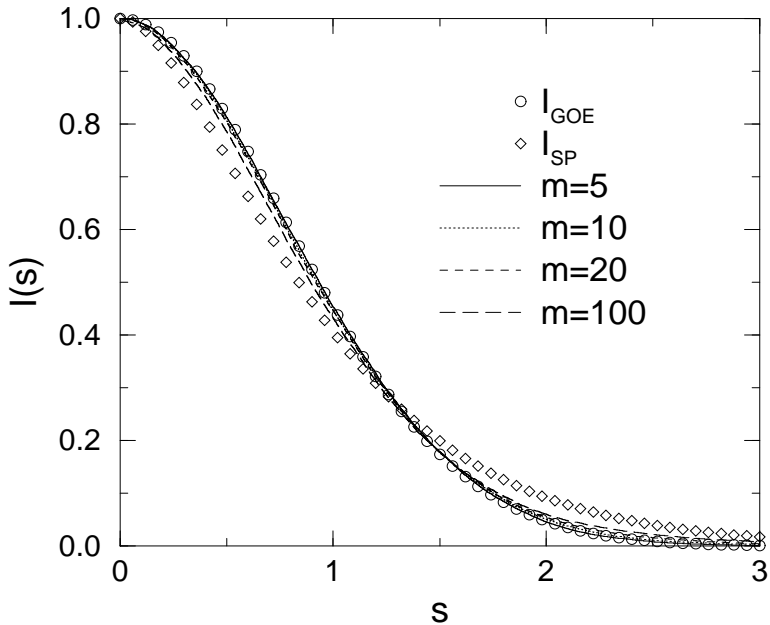,width=0.8\textwidth}}
  \caption{ILSD $I(s)$ obtained for one sector of the $D_8$-symmetric 
    patch with different parameters $m$ in the unfolding procedure.
    The result is compared to $I_{\rm GOE}(s)$ (circles) and the
    intermediate (``semi-Poisson'') statistics $I_{\rm SP}(s)$
    (diamonds).}
\label{fig4}
\end{figure}

\end{document}